\documentclass{xeus}
\usepackage{graphics}
\usepackage{psfig}
\def\asca{{\sl ASCA }}
\def\sax{{\sl BeppoSAX }}
\def\ros{{\sl ROSAT }}

\def\ein{{\sl Einstein }}

\def\xmm{{\sl XMM-Newton }}

\def\chandra{{\sl Chandra }}

\def\kmps{\hbox{km $s^{-1}$}}

\def\apj{ApJ }
\def\pasp{PASP}
\def\pasj{PASJ}
\def\aap{A$\&$A}

\def\it{\sl}

\begin{document}
\onecolumn
\title{Highly Resolved X-ray Spectra with Chandra: Dynamics of Neutral and Ionized Matter }
\author{Norbert S. Schulz}
\institute{Massachusetts Institute of Technology,
Center for Space Research, Cambridge, MA 02139, USA}
\maketitle

\begin{abstract}
 The High Energy Transmission Grating Spectrometer (HETGS) onboard the Chandra X-ray Observatory
has so far produced a large number of high resolution X-ray spectra with unprecedented
spectroscopic details. Spectra from outflows in galactic and extragalactic X-ray sources indicate
plasmas with a wide range of properties. Optically thick fluorescent matter and warm
photoionized regions play as much a role as very hot regions where collisional
ionization and scattering dominate the emission. Through the measurements of blue- and redshifts
the complex dynamics of these plasmas is revealed. Quite intriguing in this respect is the case
of X-ray absorption of neutral matter. In many cases spectral features are found to be of high
complexity though the detection of edges from intermediate Z elements 
as well as absorption lines from monatomic species to molecular compounds.
With the application of line diagnostic tools and more accurate atomic data bases we are now able to model
the properties of these plasmas as well as measure line shifts and shapes to constrain their
spatial distribution and dynamics. In various examples, i.e. plasmas from accretion disks, winds,
stationary clouds as well as the ISM, the power of highly resolved X-ray spectra is demonstrated
and the scientific capability of XEUS in this context is explored.
\end{abstract}

\section{Introduction}

With the launch of the Chandra X-ray Observatory in 1990 we entered a new era of high resolution 
X-ray spectroscopy that enriched our views throughout the entire field of high energy astrophysics.
Specifically the two grating spectrometers, the HETGS and the Low Energy Transmission Grating Spectrometer
(LETGS), with a spectral resolving power of over an order of magnitude larger than most previous
space-born X-ray spectrometers have already quite impressively demonstrated the importance and need
for high resolution instruments in the field. While focussing mostly on 
bright galactic X-ray sources, I will explore some aspects of X-ray spectroscopy and plasma diagnostics
using the HETGS which are now state of the art in the field.

X-ray binaries are the brightest X-ray sources in the sky and their bright X-ray emission is a consequence
of the accretion process of material from a close companion onto a compact object, likely
a neutron star of a black hole (BH). The result is a strong X-ray continuum that can be detected and measured
with fairly low resolution detectors like proportional or gas scintillation counters. The latter offer
spectral resolving powers of about $\Delta$E/E of 1 to 10 depending on energy. These continua have been modeled
to be of either thermal nature like blackbody radiation from the surface of a neutron star, disk blackbody
radiation from a hot inner disk surface, thermal bremsstrahlung with high energy cut-offs, power laws from
synchrotron radiation and/or reprocessed radiation through inverse Compton scattering, just to name a few. These
spectra are generally modified by continuum (photoelectric) absorption in the ISM and optically
thick matter intrinsic to the sources. Discrete emission like Fe K lines was merely detected as
local perturbations in the continuum modeling process. Lewin et al. (1995) offers a more complete review with references.
With the advent of charge coupled devices (CCDs) the resolving
power increased to the order to 10 to 60. Some good progress was made to detect discrete line emission
and absorption predominantly from faint extragalactic source like 
warm absorbers and relativistic iron lines in active galactic nuclei.
Spectra from very bright X-ray binaries offered quite little in this respect. 
Here only a few are to name like 4U 1626-67 (Angelini et al. 1995), Cyg X-3 (Kitamoto et al. 1994), and various high mass X-ray
binaries like Vela X-1, Cen X-3 and GX 301-2 (Nagase et al. 1994). The problem with X-ray binaries
was thought to be photon pile up in the CCD frame, which at such brightness levels wiped out potential line emission.

Why do we expect discrete emission in X-ray binaries in the first place? The answer is that there are many different
plasma environments in these systems that re-process the radiation of the central source. 
In the following I chose a few examples
of such environments and focus on the fact that we need various levels of spectral resolving powers to not only detect
descrete features but also resolve them. The range of parameters in these plasmas is quite large as we observe
temperatures between 10$^4$ and 10$^8$ K, densities from as low as $\sim 10^{8}$ cm$^{-3}$ in winds to 
as high as $\sim 10^{18}$ cm$^{-3}$ in accretion disks, optical depths between 0.01 and 100, ionization parameters
of up to 2$\times 10^4$ and a vast range of ionization stages. For in depths reviews I recommend Liedahl (1997) and
Paerels (1997). 
Many times conditions are complicated by the fact that
most of these plasmas are neither in ionization equillibrium nor at rest and we thus have to deal with all kinds of
plasma dynamics.      

The observation of X-rays in stars was discovered with \ein, but the fact that many stars exhibit significant X-ray
emission was finally established with the advent of \ros in 1990.
Observation and analysis of stellar spectra is difficult
and highly complex. Although most of the stars show X-ray emission at some level, but it is primarily the emission
from very young late type and early type stars or stars that are very close to the sun
that dominate the observations.
The reason is that the two former show X-ray luminosities that are many orders of magnitudes
higher than most late type main sequence stars.
On the other hand, the X-ray spectra of stars are more or less pure line spectra that need high spectral resolving power.

\section{Line Emission}

\begin{figure}
\centerline{\psfig{figure=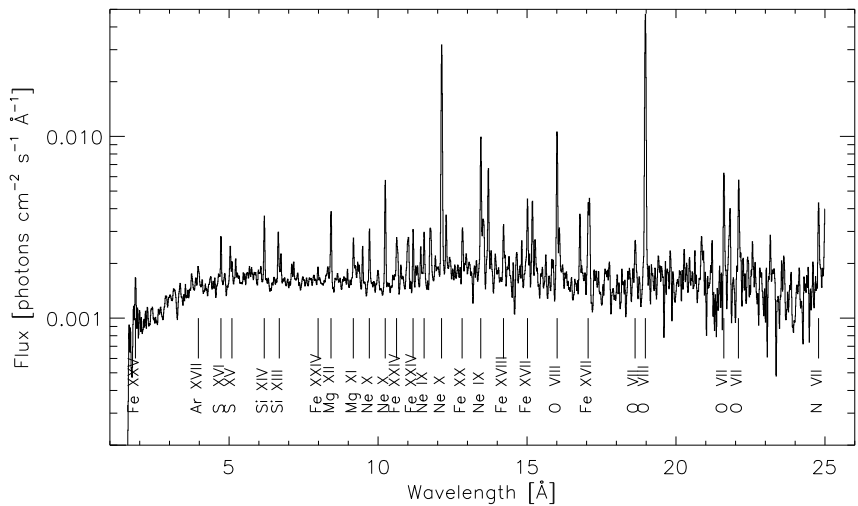,width=12.0cm}}
\caption[]{HETGS spectrum of II Pegasi (from Huenemoerder, Canizares $\&$ Schulz 2001).}
\end{figure}

Chandra so far has produced a wealth of X-ray spectra showing discrete line emission, most of these
spectra are attributed to either coronal emission or wind emission (or both) from stars. Coronal emission
in general reflects emissivity conditions in the corona of our sun, which is a hot gas of
temperatures above 10$^6$ K at relatively low densities ($\sim 10^{10}$ cm$^{-3}$). Figure 1 shows the 
example of such a spectrum observed in II Pegasi with the HETGS (Huenemoerder, Canizares $\&$ Schulz 2001). II Pegasi 
is what we call a coronally active binary and its plasma has a range of temperatures, densities, velocities, and
non-equillibrium states. In order to diagnose this plasma we usually oversimplify conditions and assume
a coronal approximation where we assume collisional excitation and ionization from the ground state and radiative
and di-electronic recombination only. Photoexitation of metastable states and photoionization are assumed to be
negligible, except for density sensitivity. The spectrum shows - besides some thermal bremsstrahlungs continuum -
a manifold of lines from most abundant ions between Z=7 (nitrogen) and Z=26 (iron). In order to determine  
emission measure distributions and abundances (in flare and non-flare states) we need
to resolve as many lines in the spectrum as possible. At a region between 10~\AA~ and 10~\AA~ we specifically
observe a large line density, which is primarily due to the fact that here we find dominant lines from Fe L shell ions. 
The mean resolving power of the Medium Energy Grating (MEG) here is 1200 with an absolute wavelength
scale of 50 to 100 km $^{-1}$ allowing for unambiguous line identifications and line flux determinations.
At lower wavelength below 10~\AA~ the resolving power of the gratings decrease, but so does the line density
in the spectrum. For example lines from H- and He-like sulfur ions can be observed at a resolving power of 500.

\begin{figure}
\centerline{\psfig{figure=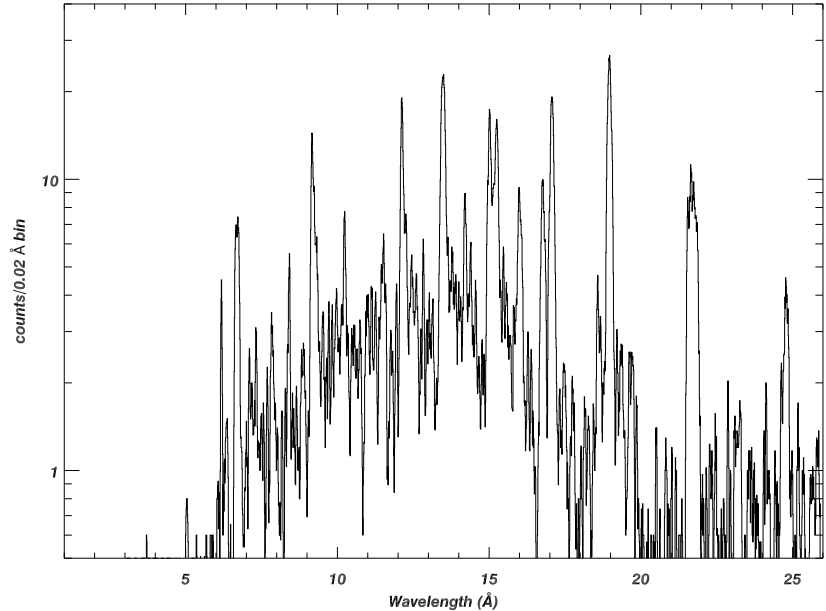,angle=90,width=12.0cm}}
\caption[]{Preliminary HETGS count spectrum from $\iota$ Ori.}
\end{figure}

Of course, this is valid only if the line are not significantly broadened and shifted due to motions in the plasma.
Figure 2 shows the example of $\iota$ Orionis, a highly eccentric early type binary (O9 III + B1 III), where we
expect lines from a high velocity stellar wind (v $\sim$ 2500 km $^{-1}$) and a collisionally ionized plasma
of a temperature of 6 10$^6$ K. Here the lines appear severely Doppler broadened, which allows us to fully
resolve the lines with a much lower power but at the expense of line confusion. The O VII triplet, for 
example well resolved in II Peg, appears as one broad line in  $\iota$ Ori although the resolving power
of the instrument is the same. The reason is that in II Peg the line are generated in coronal loops near the
stellar surface, while in $\iota$ Ori the lines are generated in the wind 
a few stellar radii away from the surface of the star 
where the terminal velocity is reached. 

\begin{figure}
\centerline{\psfig{figure=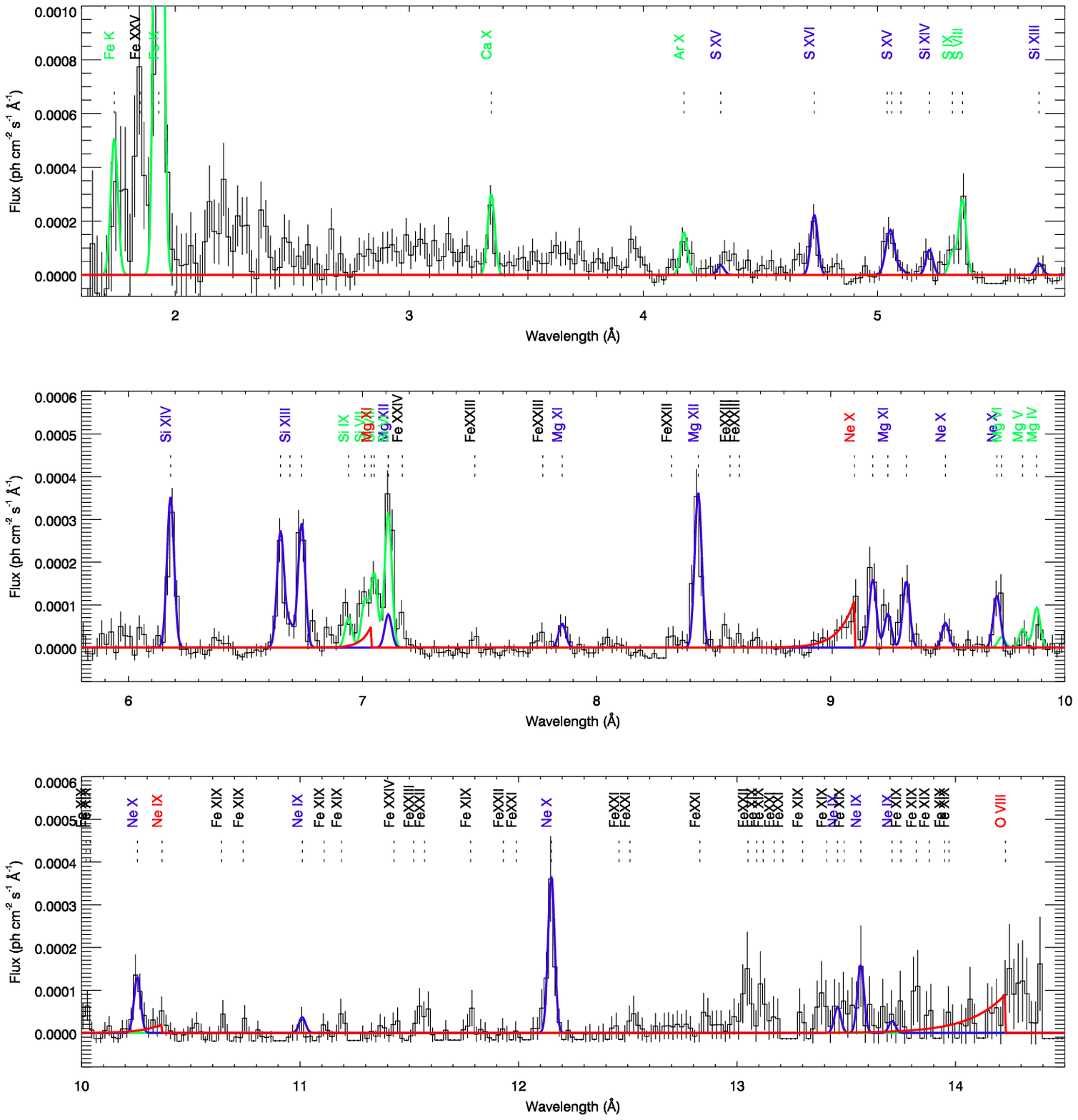,width=12.0cm}}
\caption[]{The residual line spectrum of Vela X-1 during eclipse observed with the HETGS.}
\end{figure}

When we observe line emission from accreting X-ray binaries we primarily find that plasmas are predominantly 
photo-ionized, at most hybrid plasmas (Bautista and Kallman 2000, Porquet and Dubois 2000). 
Exceptions from this are rare, like high temperature outflow in the jet of SS 433 (Marshall, Canizares, $\&$
Schulz 2002) or possible contributions from shock instabilities in the stellar wind of 
high mass X-ray binary (HMXBs) companions as was speculated in Cyg X-1 (Schulz et al. 2002), 
just to name a few. 

Reprocessed X-rays in a stellar winds result in discrete emission lines. Figure 3 shows an example
of an HETGS spectrum of the ionized stellar wind in Vela X-1 during eclipse (see Schulz et al. 2002).
The spectrum shown is a residual spectrum, i.e. the scattered continuum from the central
source has been subtracted and it thus appears entirely resolved into emission lines. The 
lines originate from at least two dynamically different regions. First, we observe lines 
from the photo-ionized wind (blue). The lines itself are actually not resolved, i.e. we
do not observe any measurable Doppler-broadening above the resolution limit, which at Ne X
corresponds to about 250 \kmps. There are also no line shifts. A detailed analysis of the
spectrum reveals signatures of photo-excitation in the plasma (Schulz et al. 2002).
This is most prominently visible in the He-like triplets as the resonance lines
appear much too strong as expected from a photo-ionized or hybrid plasma (Porquet $\&$ Dubois
2000). Wojdowsky et al. (2002) in an analysis of a similar spectrum of Cen X-3 interprets
this effect as resonant scattering of the source continuum in the resonance line. 
We also observe radiative
recombination continua (RRCs, red) from the photo-ionized gas from Mg XII, Ne X, and O VIII ions, although
the one from Mg XII is blended in with another line complex. RRCs
appear narrow at low temperatures and broad at high temperatures and it thus critically depends
on the resolution of the instrument how accurately we can measure the temperature of the 
photoionized gas. In the HETGS
spectrum they appear fully resolved at a temperature of 1.2 10$^5$ K. The most remarkable
lines in the spectrum in Figure 3, however, are the fluorescence lines, specifically the 
ones at S, Si, and Mg. Here the observation of some L-shell ions is most likely an indicator of
cold material of high column density. 
The challenge we face here from the point of spectroscopy is that low excitation fluorescence lines
(House 1969) cannot be resolved with the HETGS resolution and the determination of the ion type
becomes strongly model dependent.    

\begin{figure}
\centerline{\psfig{figure=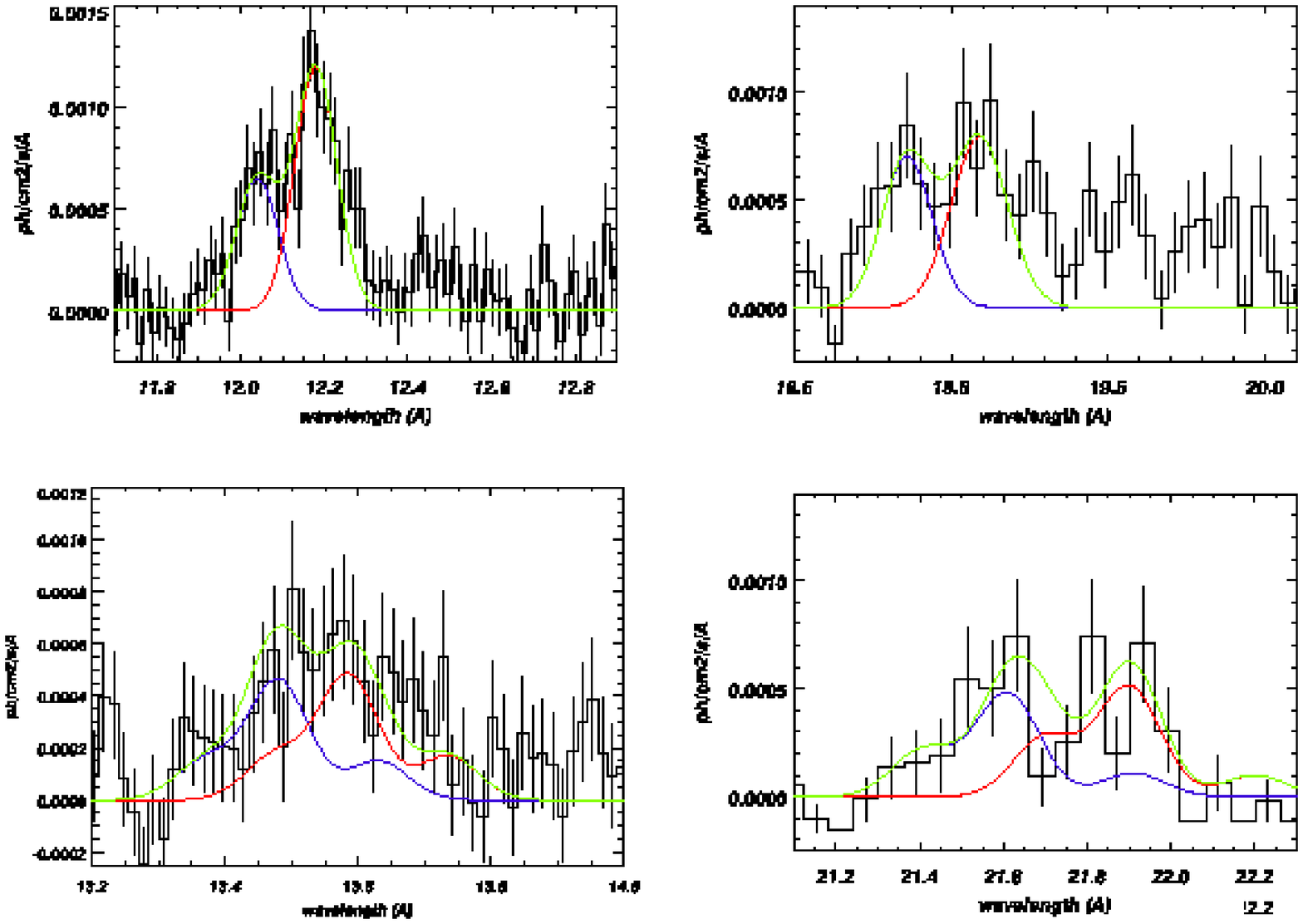,width=12.0cm}}
\caption[]{Doppler line pairs from the accretion disk in 4U 1626-67 (from Schulz et al. 2002a)}
\end{figure}

Emission lines are also observed with the HEGTS from accreting disks in low-mass X-ray binaries 
(Brandt $\&$ Schulz 2001, Cottam et al. 2001, Schulz et al. 2002a). In 4U 1626-67 it was well
known from previous \asca and \sax observations that the spectrum shows strong Ne and O lines
(Angelini et al. 1995, Owens, Oosterbroek, $\&$ Parmar 1997).
The HETGS spectrum basically confirms these observations as only the 
lines from H-like and He-like ions of Ne and O were detected. Surprisingly the lines
appeared fully resolved and extremely broad, which turns out to be an effect of blue-
and red-shifted line components as shown in Figure 4. At velocities between 1550 to 2610 
\kmps for the blueshifted and 770 to 1900 \kmps for the redshifted line components, spectral
resolution does not seem to be that crucial and in fact only a factor of resolution higher
than CCD resolution would have sufficed to detect the Doppler split. However, it becomes
critical again when we want to model the He-like triplets, which now take on a
very complex shape in which the resonance, intercombination, and forbidden line components
of the blueshifted line triplet now mix with the redshifted triplet (see bottom of Figure 4).     
   
\begin{figure}
\centerline{\psfig{figure=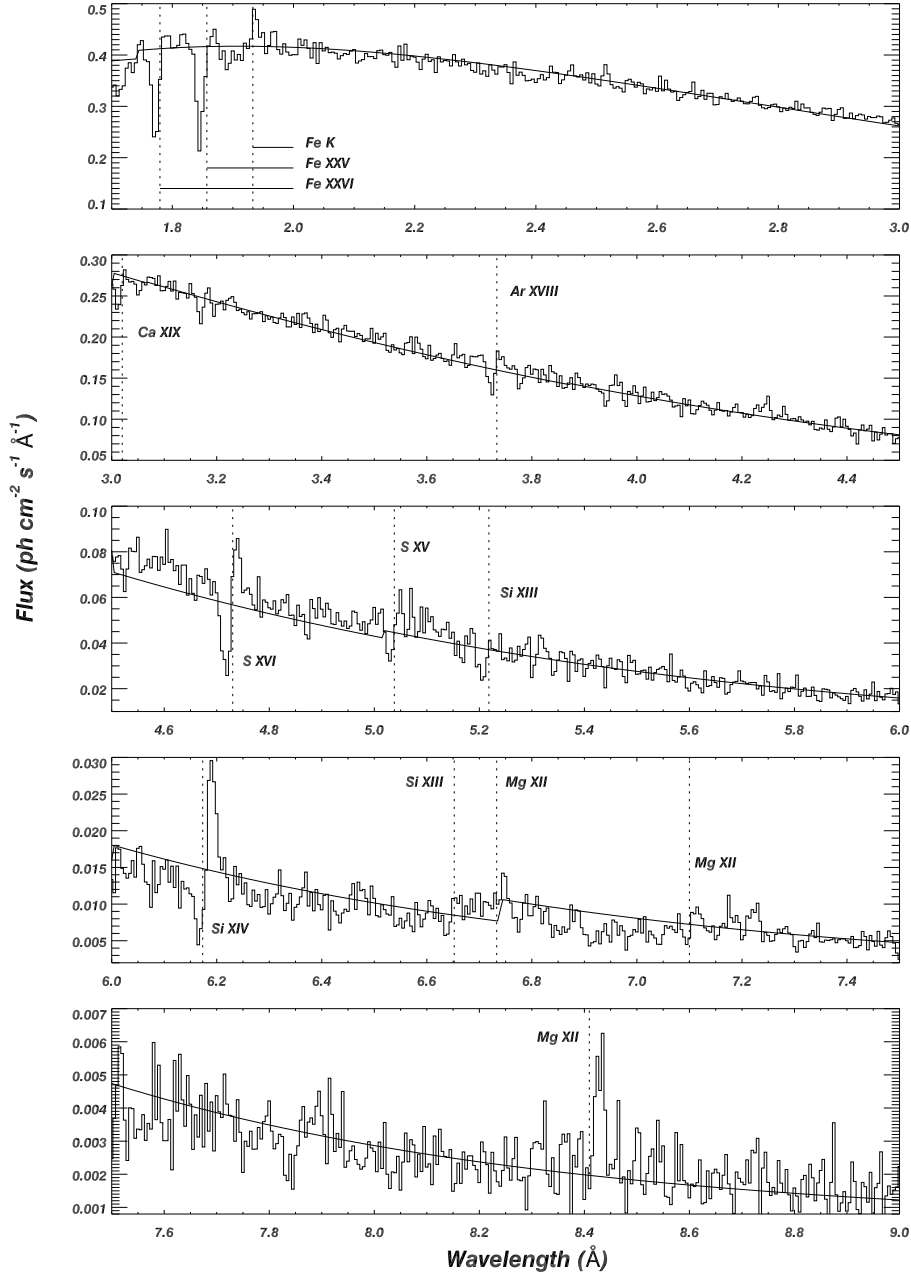,width=12.0cm}}
\caption[]{HETGS spectrum from a deep intensity dip of Cir X-1 during periastron (from Schulz $\&$ Brandt 2002)}
\end{figure}

\section{Resonant Absorption}

Besides line emission we on many occasions observe resonant line absorption. Most illustrative
in this respect is the case of Cir X-1, in which we found P Cygni line profiles for the
first time in X-rays (Brandt $\&$ Schulz 2000). Figure 5 shows the spectrum of Cir X-1 during
a major dip, i.e. when major parts of the spectrum appear heavily absorbed (from
Schulz $\&$ Brandt 2002). Besides
strong emission lines from all H- and He-like ions of abundand elements we observe strong
and blueshifted absorption acompaning these lines. Brandt $\&$ Schulz (2000) interpret
these absorption lines as a consequence of a fast outflow/wind originating at the 
line production site resulting in P Cygni lines. These lines are 
narrow with blueshifts ranging from 400 to 2000 \kmps. With medium resolution detectors
like \asca these lines were not seen and only barely detected with the XMM-RGS
(Shirey 2001, priv. comm). These absorption lines appear variable
on timescales of one to a few hours (Schulz $\&$ Brandt 2001). A detailed analysis
revealed that this variability is most likely the consequence of fluctuations
in the ionization fraction of the outflow (Schulz and Brandt 2002). A result like this 
could not have been possible without the superb spectral resolution of the instrument.

Other examples includes HETGS spectra of Cyg X-1. Here the situation appears similarly
complex. We observed this black hole binary at various orbital phases and find 
a series of weak and narrow absorption lines (Schulz et al. 2002b, Marshall et al. 2001,
Miller at al. 2002). Parts of HETGS spectra from two phases are 
illustrated in Figure 6. The top two panels show spectra from an orbital phase where
we view the focussed stellar wind of the star with a significant velocity component that points
away from the observer. As a consequence we observe red-shifted absorption lines (Marshall
et al. 2001) by about 450 \kmps. The bottom panel in Figure 6 shows an orbital phase, where there
is no such velocity component and here we see absorption lines without 
measureable red-shifts (Miller et al. 2002). I have to point out again that 
the detection of these dynamic lines require the maximum resolution of the HETGS device
(in  1st order).       

\begin{figure}
\centerline{\psfig{figure=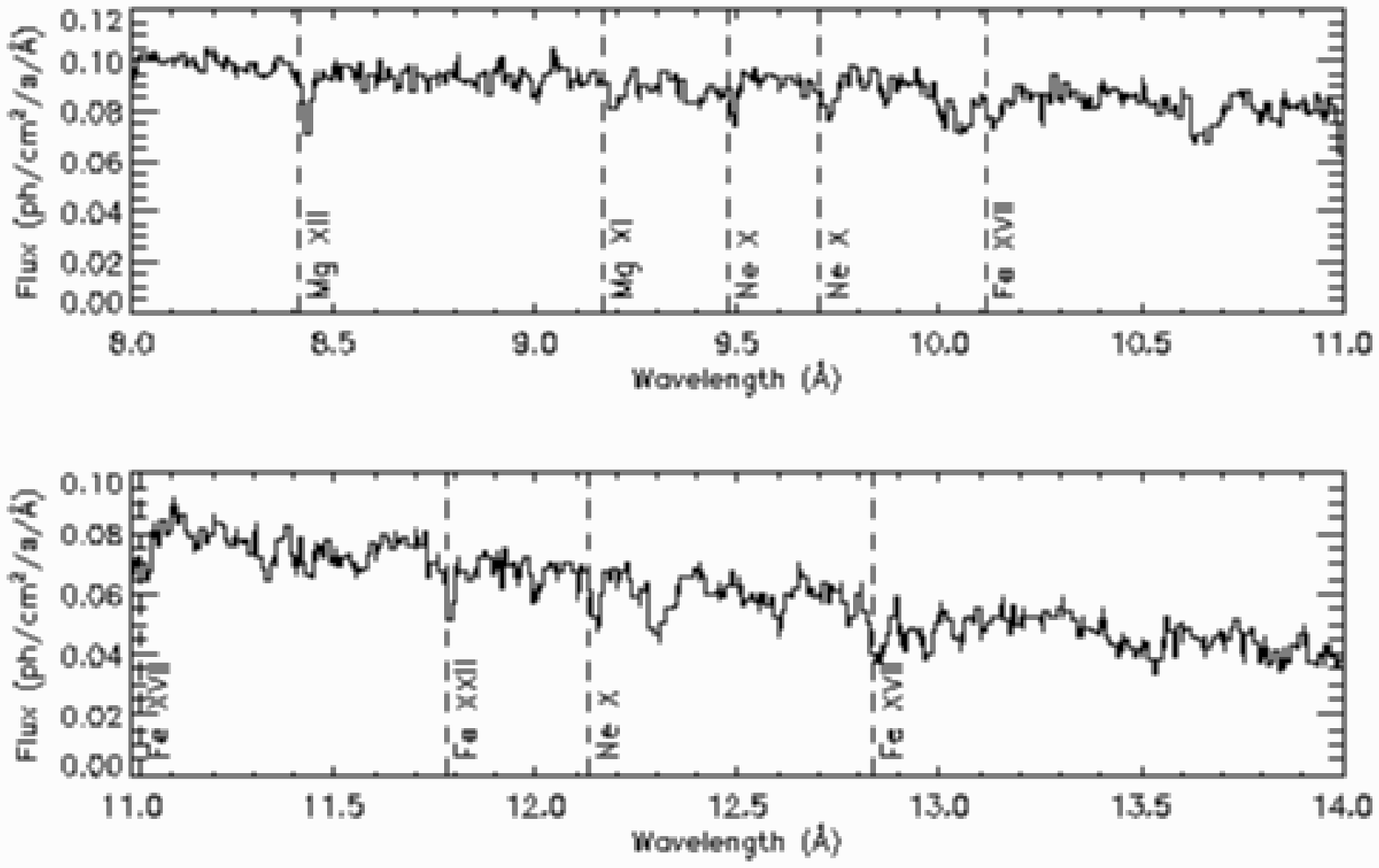,width=12.0cm}}
\centerline{\psfig{figure=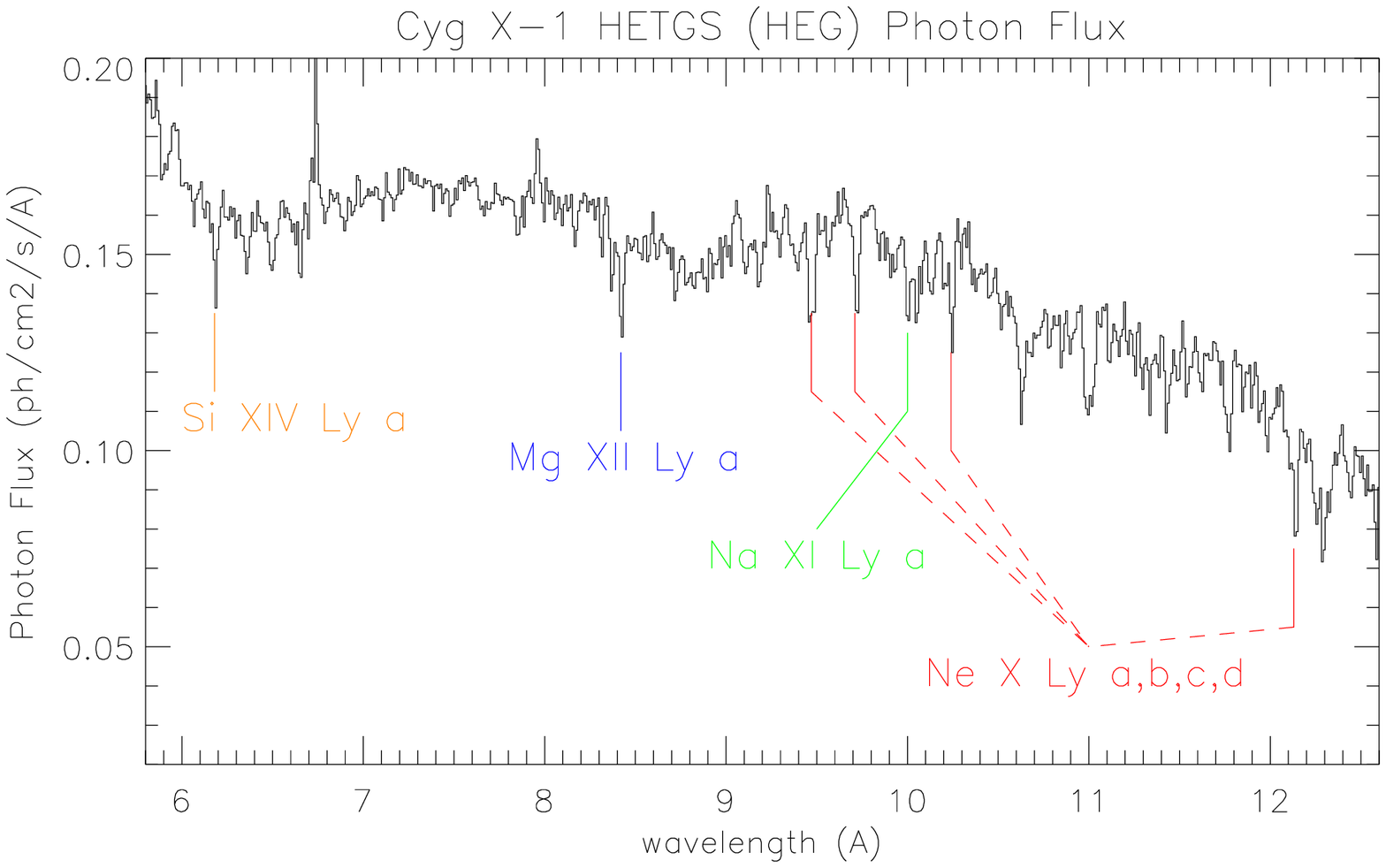,width=12.0cm}}
\caption[]{Top: redshifted absorption lines in the spectrum of Cyg X-1 at orbital phase 0.82 (Marshall et al. 2001).
Bottom: Unschifted lines at phase 0.74 (Miller at al. 2002).}
\end{figure}

\section{Photoelectric Absorption}

The understanding of photo-electric absorption of X-rays is of immediate importance for the
analysis of X-ray spectra. With the high resolution X-ray spectrometers onboard \chandra and \xmm~
we can now directly measure the depth and structure of major photo-electric edges.
The treatment of photo-electric absorption has a longstanding history in determining 
energy dependent photoionization cross-sections of the interstellar medium ISM from the far UV to
X-rays (Strom $\&$ Strom 1961, Brown $\&$ Gould 1970, Morrison and McCammon 1983). 
Most recently Wilms, Allen $\&$ Mc Cray (2000) presented the latest improvements
on the abundance distribution of the ISM. We have started a survey of observations with the HETGS
of bright X-ray binaries in the galactic plane in order to determine the optical depths
of low energy absorption edges from Ne, Fe, O, and possibly N. In two pilot studies
by Paerels et al. (2000) and Schulz et al. (2002) several remarkable features have been
found ranging from strong absorption of the 1s-2p atomic absorption line from oxygen as well
as indications of narrow absorption from various oxides.

Figure 7 shows three examples of measured Fe L edges observed in X-ray binaries with
different interstellar column densities. Of course, in general we cannot easily
distinguish between source intrinsic and ISM contributions and here we have an opportunity
to advance our understanding of cold matter structure. The spectra around 710 eV show
that the edge shows considerable structure, mainly a three-fold absorption pattern 
corresponding to a very weak Fe L1 (844 eV, not included in Figure 7) and strong
Fe L2 (719 eV) and Fe L3 (706 eV). The values here are given for metallic Fe. The Fe L edge structure
from metallic iron was measured recently by Kortright $\&$ Kim (2000) and we used their
cross sections to fit the data (straight lines in Figure 7). Clearly, as this edge
shows significant structure over a large range of column densities (here
1 to 9 10$^{21}$ cm$^{-2}$) it probably caused residuals in spectral fits with 
low resolution detectors which ultimately were interpreted as low energy Fe lines.
However, even if we include the basic structure of this edge, the three examples
in Figure 7 show that there is more detailed structure accompanying these edges.
Here we see great challenges for the future. 
 
\begin{figure}
\centerline{\psfig{figure=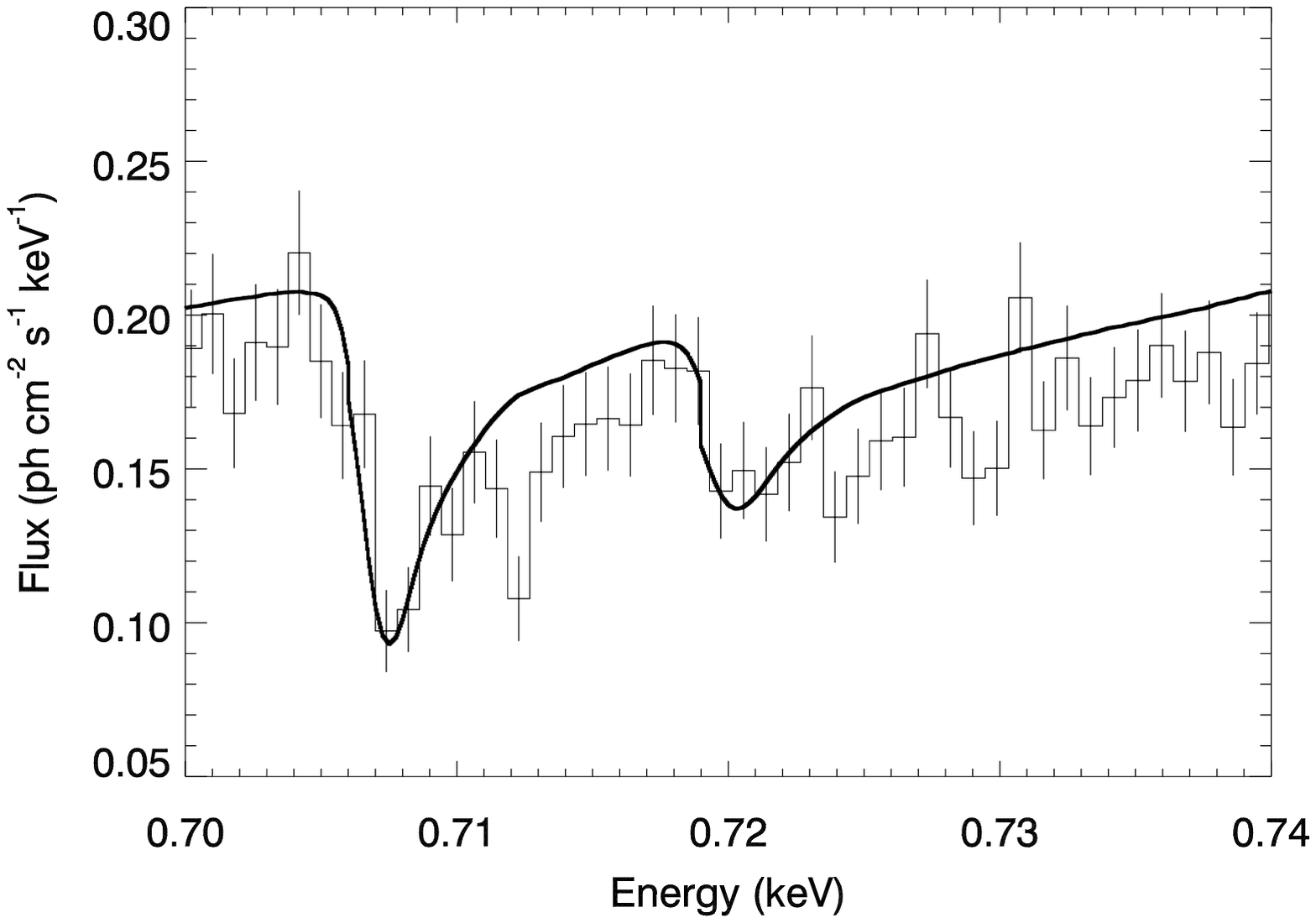,width=6.0cm}\psfig{figure=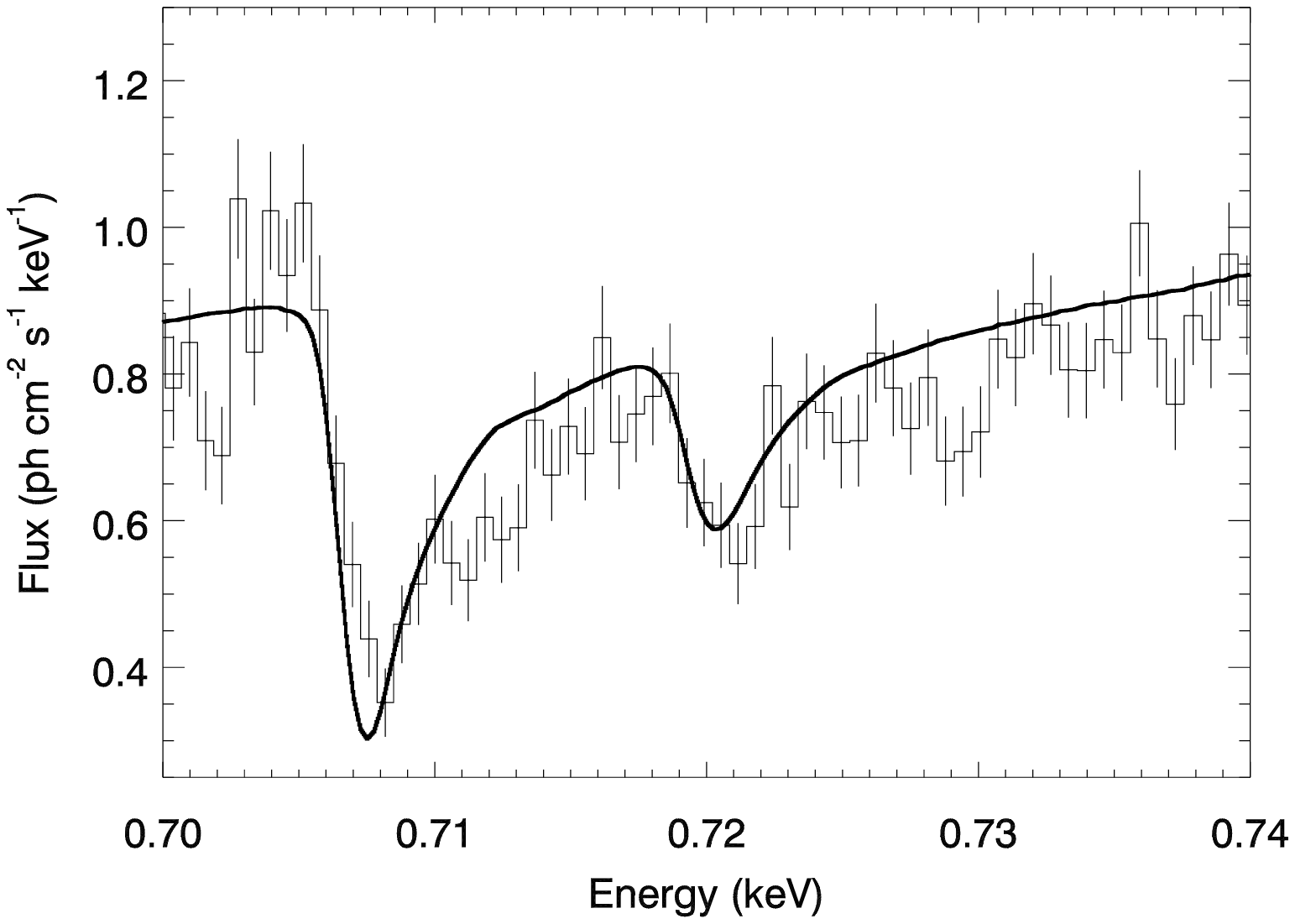,width=6.0cm}\psfig{figure=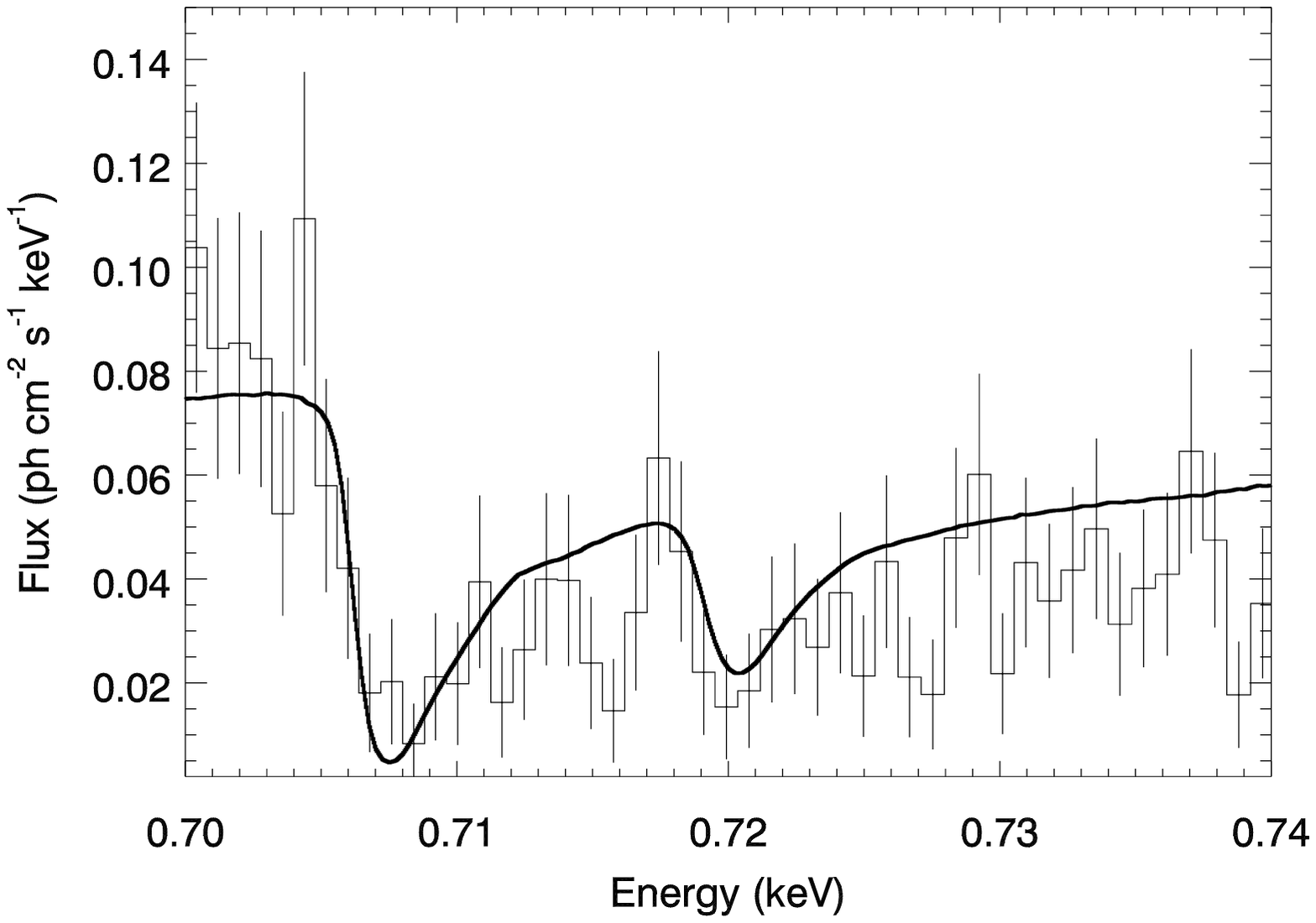,width=6.0cm}}
\caption[]{Three examples of Fe L 2 and 3 edge observations at various column densities: 4U1636-53 at 3 10$^{21}$ cm$^{-2}$ (left),
Cyg X-1 at 6 10$^{21}$ cm$^{-2}$ (middle), and GX 349+2 at 9 10$^{21}$ cm$^{-2}$ (right)}
\end{figure}

\section{Scattering}

This topic is very much related to the previous one as we now enter an era of solid state
astrophysics in the X-ray domain. 
The understanding of absorption and scattering of X-rays is not only of immediate importance for the
analysis of X-ray spectra, but also fundamentally contributes to our understanding of the composition
and structure of the ISM and the circumstellar medium. X-ray absorption fine structure (XAFS)
are produced by the interference effect of a back-scattered electron wave from nearby atoms 
on the final state of the out-going electron wave. XAFS thus depend on the type, structure, 
and state of the intervening matter. It now allows us to study
the properties of interstellar grains and molecules in the 
spectra that have passed through interstellar matter (Woo 1995).

Figure 8 shows two examples where we likely observe solid matter signatures from
circumstellar matter (left) and ISM matter (right). In HETGS spectra of GRS 1915+105,
Lee et al. (2002) found XAFs signatures near the Si edge, but also indications
near S and Mg. This is quite important in the context of 
observations of clouds and dust in the IR and UV as well as from meteorites and
interplanetary dust, where it is generally believed that interstellar grains are 
composed of various combinations of ice (H$_2$O), graphite (C), silicates 
(e.g. FeSiO$_3$, MgSiO$_3$), magnetites (e.g. Fe$_3$O$_4$), to name a few. From the
instrumental point of view, two factors are important here: spectral resolution \it and \rm
effective area as most of these features are extremely weak. GX 5-1 is exceptionally
bright even though it lies in the galactic bulge at a projected distance of 10 kpc
and is thus strongly affected by absorption. The right part of
Figure 8 shows a strong edge signal in the count spectrum at Si K. Once the 
instrumental contribution, which is largely affected by contributions from
SiO$_2$, is removed we still observe large residuals of SiO$_2$. The signal
is significantly stronger than what would be expected from calibration uncertainties
and it is quite likely that we observe interstellar SiO$_2$.   

\begin{figure}
\centerline{\psfig{figure=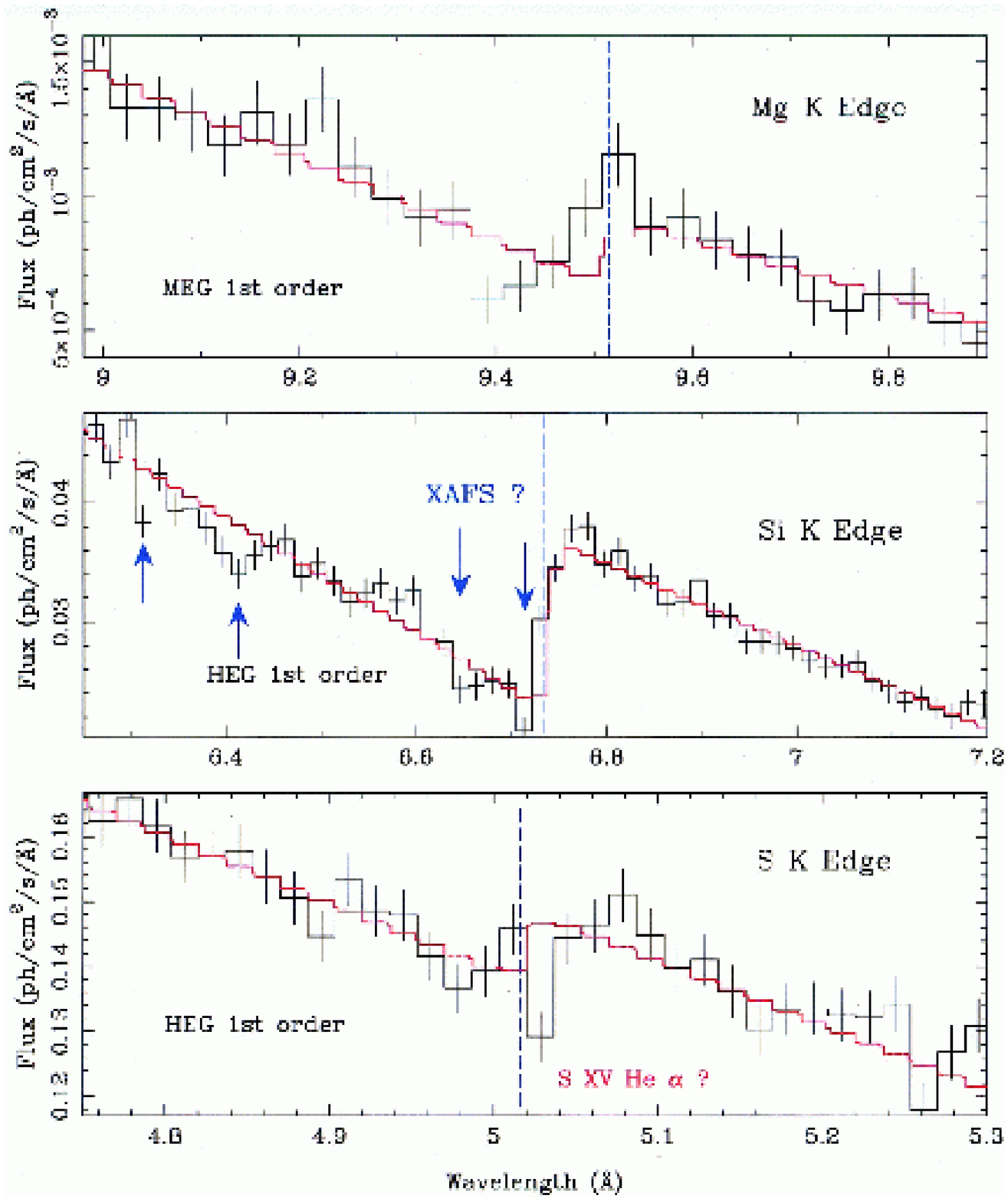,width=8.0cm}\psfig{figure=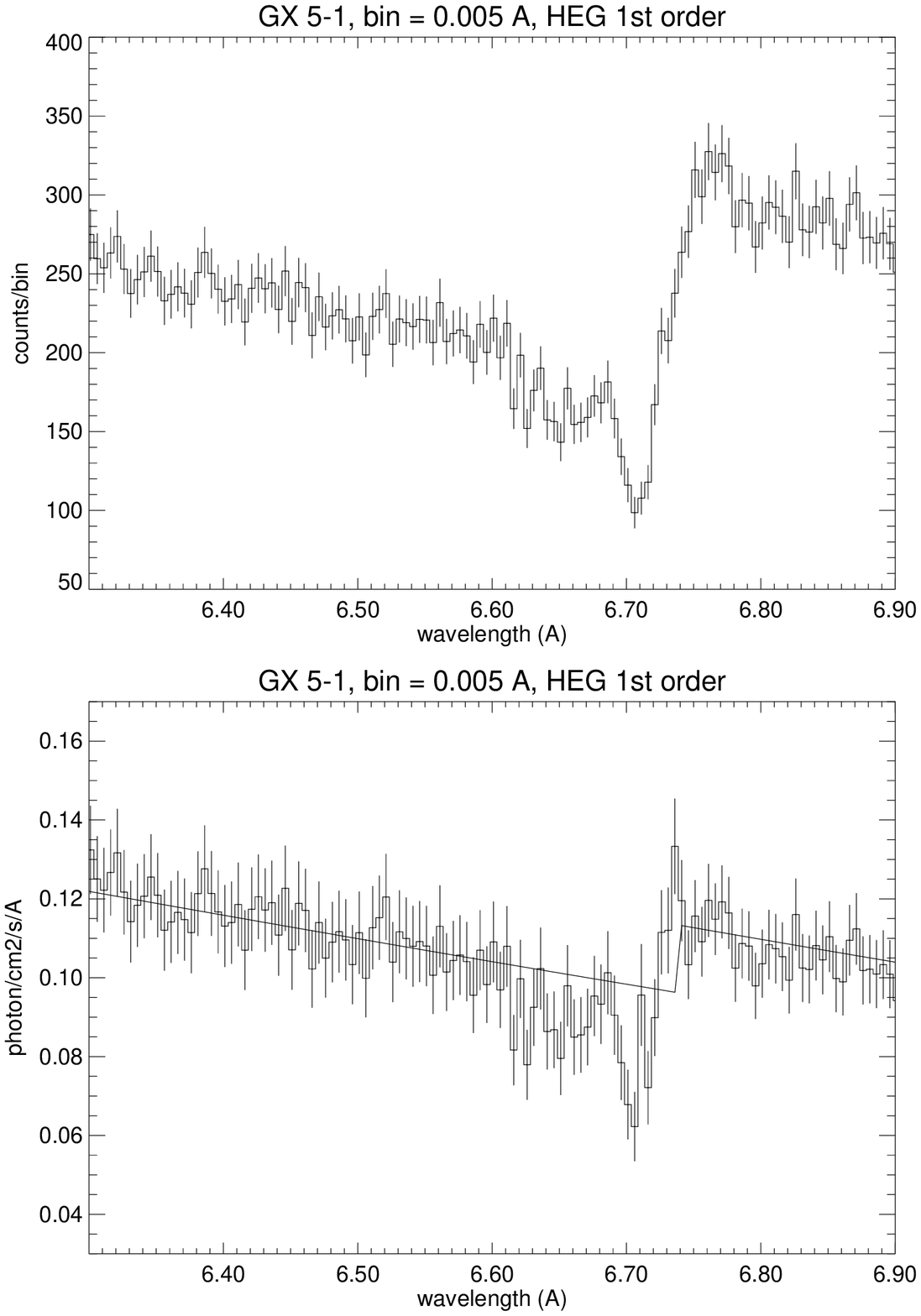,width=7.0cm}}
\caption[]{Left: XAFS near edges in GRS 1915+105 (from Lee et al. 20012.
Right: Strong residual structures in the Si K edge of GX 5-1.}
\end{figure}

\section{Implications for XEUS}

These last sections clearly demonstrated that highly resolved spectroscopy opened a new window
to our galaxy in X-rays. A common thread in these new discoveries is that they were clearly not 
possible without the unprecedented spectral resolving power of the HETGS. XEUS, with specification
of 2 eV (narrow field of view) in the range of 0.05 to 30 KeV (250 to 0.4 ~\AA~), offers
not only a continuation of these capabilities, but will extend our currently accessible
bandpass. Here are a few science topics, which I can identify in this respect:

\begin{itemize}
      \item line shape studies - Doppler tomography in accretion disk lines
      \item line variability studies - reverberation mapping
      \item weak edge structures - cold matter in X-ray sources and the ISM
      \item XAFS in the ISM - grains and molecules
      \item ionized absorption in the IGM
      \item extended source - abundance, temperature, velocities in SNRs
\end{itemize}

As the observation of fainter sources with the HETGS and \xmm-RGS 
is quite exposure time
consuming and the study of line variability is exposure limited, its so much larger effective area
will clearly push our current limits. At higher energies the extended bandpass will allow
to include the observation of the full Fe XXIV Lyman $\alpha$ series, H- and He-like Ni ions, as well
as highly resolved synchrotron absorption lines. The increased resolving power at high energies
is specifically interesting as we will be able to better resolve narrow Fe K fluoresecence lines, which
are already found in quite a variety of sources. As for the continuation of our current efforts,
however, it also has to be concluded that a resolving power of 1 eV at 1 keV, which is currently
only listed as a goal for the mission, is clearly more desired than the current specification of 2 eV.
The latter would severely limit our diagnostical potential for spectroscopic features similar to the ones 
shown in this presentation.

\begin{acknowledgements}
Work at MIT was supported by NASA through the HETG contract NAS 8-38249 and through the Smithonian
Astrophysical Observatory (SAO) contract SVI-61010 for the Chandra X-ray Center. In addition I want
to thank Raquel Morales for critically reading through the manuscript as well as the CXC and HETG team
at MIT for their continuing support. 
\end{acknowledgements}

\end{document}